\documentclass[prl,aps,twocolumn,showpacs,reprint,amsmath,amssymb,floatfix]{revtex4-2}
\usepackage{amsmath}
\usepackage{graphicx}
\usepackage{dcolumn}
\usepackage{bm}
\usepackage{float}
\usepackage{xcolor}
\usepackage{lineno}
\usepackage{subfigure}
\usepackage{ulem}
\usepackage[colorlinks=true,linkcolor=blue]{hyperref}
\usepackage{silence}
\UseRawInputEncoding
\begin{document}

\preprint{APS/123-QED}

\title{Origin of Bright Quantum Emissions with High Debye-Waller factor in Silicon Nitride}

\author{Shibu Meher}
\author{Manoj Dey}
\author{Abhishek Kumar Singh}
\affiliation{%
Materials Research Centre, Indian Institute of Science, Bangalore 560012, India}%

\date{\today}

\begin{abstract}
Silicon nitride has emerged as a promising photonic platform for integrated single-photon sources, yet the microscopic origin of the recently observed bright quantum emissions remains unclear. Using hybrid density functional theory, we show that the negatively charged N$_\text{Si}$V$_\text{N}$ center (NV$^{-}$) in the C$_{1h}$ configuration exhibits a linearly polarized zero-phonon line (ZPL) at 2.46~eV, with a radiative lifetime of 9.01~ns and a high Debye-Waller (DW) factor of 33\%. We further find that the C$_{1h}$ configuration is prone to a pseudo-Jahn-Teller distortion, yielding two symmetrically equivalent defect structures that emit bright, linearly polarized ZPL at 1.80~eV with a lifetime of 10.17~ns and an increased DW factor of 41\%. These nitrogen-vacancy-related  defects explain the origins of visible quantum emissions, paving the way for deterministic and monolithically integrated silicon-nitride quantum photonics.

\end{abstract}

\maketitle

Bright single-photon emitters operating near the visible spectrum have attracted significant interest, exemplified by the $\sim$2~eV emitter in hBN~\cite{Aharonovich2016} and the NV$^-$ center in diamond~\cite{Doherty2013}. Hybrid integration approaches, which combine different materials for the emission source and photonic elements, increase fabrication complexity and introduce coupling losses at interfaces~\cite{Aharonovich2016,Martin2023}. A unified photonic platform with built-in single-photon emitters is therefore crucial for achieving direct, scalable, and low-loss integration of quantum light sources. Silicon nitride has emerged as a state-of-the-art quantum photonic platform, offering a wide bandgap suitable for hosting quantum emitters. Recently, room-temperature single-photon emissions near $\sim$2~eV have been observed in silicon nitride~\cite{Senichev2021,Martin2023}. These emitters are bright, stable, linearly polarized, and exhibit high purity. Most importantly, monolithic integration of these emitters within the silicon nitride photonic platform has been demonstrated~\cite{Senichev2022}. However, the microscopic origin of these emitters remains unidentified.

Senichev \textit{et al.}~\cite{Senichev2021} reported intrinsic single-photon emitters in silicon nitride, emitting in the range of 568-722 nm (2.18-1.72 eV) at room temperature. Using a similar synthesis procedure, Martin \textit{et al.}~\cite{Martin2023} observed several emitters with ZPL emission around 548 nm (2.26 eV) with a 7 nm deviation at 4.2 K. In both studies, silicon nitride films were grown under nitrogen-rich conditions on SiO$_2$/Si substrates. Analysis of 130 emitters revealed an average second-order autocorrelation value of 0.2 at zero delay, confirming high single-photon purity. Fluorescence lifetime measurements under pulsed excitation for 44 emitters yielded lifetimes ranging from 1.4 to 3.9 ns, with an average of 2.4 ns. The emitters exhibited bright and stable emission without noticeable blinking or bleaching for up to 100 s at saturation power. More recently, Chen \textit{et al.}~\cite{Chen2025} identified single-photon emitters with emission between 570 and 582 nm (2.18-2.13 eV), showing large Debye-Waller factors (50-74\%) and lifetimes of 3-8 ns. These emitters were attributed to defects in the silica substrate, despite being synthesized using the same method as in Refs.~\cite{Senichev2021, Martin2023}. Given these consistent optical signatures yet differing interpretations, identifying the microscopic nature of these quantum emitters is essential for enabling targeted quantum-optics measurements and deterministic defect creation in integrated Si$_3$N$_4$ photonic platforms.

In this letter, we identify the microscopic origin of the $\sim$2 eV single-photon emitters observed in silicon nitride. Using hybrid density-functional theory, we investigate the thermodynamic stability, electronic structure, and optical properties of the nitrogen-vacancy-antisite (N$_\text{Si}$V$_\text{N}$, or NV center) defect in $\beta$-Si$_3$N$_4$. We show that this defect exhibits a bright, linearly polarized optical emission whose ZPL, radiative lifetime, and DW factor are in close agreement with experimentally reported emitter characteristics. We further reveal that a pseudo-Jahn-Teller distortion creates additional symmetry-broken configurations that remain optically active and fall within the same spectral window.

$\beta$-Si$_3$N$_4$ crystallizes in the centrosymmetric P6$_3$/m space group~\cite{Goodman1980}, with two inequivalent nitrogen sites (C$_{3h}$ and $m$) and a single silicon site of $m$ symmetry. In this work, we consider the nitrogen vacancy at the C$_{3h}$ site (V$_\text{N}$) and the associated nitrogen-antisite-vacancy complex (N$_\text{Si}$V$_\text{N}$). Defects are modeled in a 280-atom hexagonal supercell with $\Gamma$-point sampling of the brillouin zone. Calculations are performed using the hybrid Heyd-Scuseria-Ernzerhof (HSE) functional~\cite{heyd2003hybrid} and projector augmented-wave potentials~\cite{blochl1994projector}, as implemented in \texttt{VASP}~\cite{kresse1996efficient}. A Hartree-Fock exchange fraction of $\alpha=0.30$ yields a band gap of 6.1 eV, in good agreement with experiment~\cite{Asano2022} [see Sec. S1 of the Supplementary Information (SI); see also Refs. ~\cite{Asano2022,Zhang1991,Kumagai2014,Kavanagh2024,HR1950,Alkauskas2014,defectpl,Lax1952,Kubo1955,qijing} therein]. Unit-cell optimizations used a 4$\times$4$\times$9 Monkhorst-Pack \textbf{k}-mesh~\cite{Monkhorst1976} and a 400 eV plane-wave cutoff, giving relaxed lattice constants $a=b=7.61$ $\text{\AA{}}$ and $c=2.91$ $\text{\AA{}}$, consistent with Ref.~\cite{HARDIE1957}. Furthermore, the calculated formation enthalpy of -8.96 eV per formula unit is in good agreement with that of the experimentally reported value of -8.58 eV~\cite{OHARE1999303}. Supercell geometries were relaxed using PBE functional~\cite{Perdew1996} until forces were below 0.001 eV/$\text{\AA{}}$, followed by electronic structure calculation using hybrid-functional. Excited-state energies and geometries were obtained using the $\Delta$SCF method~\cite{Gali2009} and spin polarization is explicitly considered in all the calculations. Electron-phonon coupling was analyzed using \texttt{PHONOPY}~\cite{Togo2015} and \texttt{DEFECTPL}~\cite{Dey2025,defectpl,Meher2025}. Finite-size corrections for charged defects were applied using the eFNV scheme~\cite{Kumagai2014} via \texttt{DOPED}~\cite{Kavanagh2024}. Further computational details are provided in Sec. S2 of SI.

\begin{figure}
    \centering
    \includegraphics[width=\linewidth]{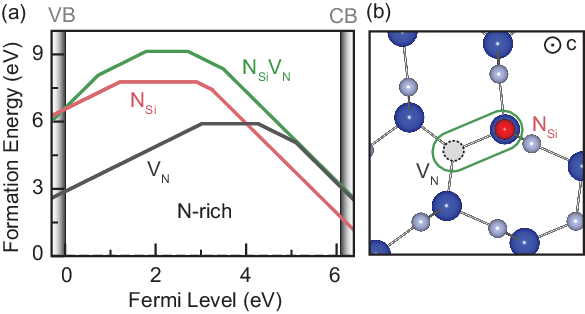}
    \caption{(a) Formation energy of nitrogen vacancy (V$_\text{N}$, grey), nitrogen antisite (N$_\text{Si}$, red), and nitrogen antisite-vacancy complex (N$_\text{Si}$V$_\text{N}$, green) under N-rich growth conditions. The slopes and kinks in the plot represent the charge states and charge transition levels, respectively. (b) Geometry of the N$_\text{Si}$V$_\text{N}$ complex, shown along the \textit{c}-axis projection. Blue, silver, and red atoms represent silicon, nitrogen, and nitrogen-antisite, respectively.}
    \label{figure1}
\end{figure}

From defect formation energy calculations, we find that the nitrogen vacancy (V$_\text{N}$) is the most stable intrinsic defect in $\beta$-Si$_3$N$_4$, consistent with previous computational and experimental studies~\cite{Grillo2011, Lenahan1990}. As shown in Fig.~\ref{figure1}(a) and Fig.~S1 in the SI, V$_\text{N}$ exhibits the lowest formation energy under both nitrogen-rich and silicon-rich conditions. The nitrogen-rich growth environments employed in recent experiments~\cite{Senichev2021, Martin2023, Chen2025} also favor the formation of stable N$_\text{Si}$ defects. Furthermore, thermal annealing used to activate the emitters~\cite{Senichev2021} can promote migration of N$_\text{Si}$ and V$_\text{N}$, enabling NV center formation throughout the Si$_3$N$_4$ matrix. Our nudged elastic band calculations suggest that the formation of the N$_\text{Si}$V$_\text{N}^-$ defect from V$_\text{Si}^-$ involves a high migration barrier of 2.66\,eV, as shown in Fig.~S4 of the SI. Unlike the NV center in diamond, which constitutes an extrinsic substitutional nitrogen adjacent to a carbon vacancy, the NV defect in $\beta$-Si$_3$N$_4$ originates from an intrinsic nitrogen antisite (N$_\text{Si}$) adjacent to V$_\text{N}$ [Fig.~\ref{figure1}(b)]. As shown in Fig.~\ref{figure1}(a), the NV center is stable in charge states $+$2 to $-$2 within the bandgap as the Fermi level varies between the valence band maximum (VBM) and conduction band minimum (CBM). The neutral NV center (NV$^0$) has a singlet ground state, whereas the charged NV$^{+}$ and NV$^{-}$ defects exhibit doublet ground states, each hosting a single unpaired electron. In contrast, the V$_\text{N}^0$ is paramagnetic with one unpaired electron, while its positive and negative charge states exhibit singlet ground states.

The relaxed geometry of V$_\text{N}$ exhibits C$_{3h}$ point-group symmetry, in contrast to the C$_{3v}$ symmetry reported by Nanataki \textit{et al.}~\cite{Nanataki2022}. In the neutral charge state, the defect hosts a single unpaired electron occupying a pair of nearly degenerate majority-spin orbitals, which drives a strong Jahn-Teller (JT) distortion of the local environment. In the high-symmetry configuration, the three neighboring Si atoms form an almost perfect equilateral triangle with C$_{3h}$ symmetry. Upon JT distortion, however, this triangle becomes scalene, and the symmetry is lowered to C$_{1h}$. The resulting symmetry breaking lifts the degeneracy of the defect orbitals and produces a large splitting of approximately 2.9 eV (see Fig. S2 in the SI). In contrast, the positively and negatively charged V$_\text{N}$ states do not exhibit a JT instability and therefore preserve C$_{3h}$ symmetry. When a nitrogen atom substitutes a neighboring Si site, forming the N$_\text{Si}$V$_\text{N}$ complex, structural relaxation yields a C$_{1h}$ configuration. This structure contains a mirror plane passing through the substituted N atom and the two adjacent Si atoms.

\begin{figure*}[t]
    \centering
    \includegraphics[width=\linewidth]{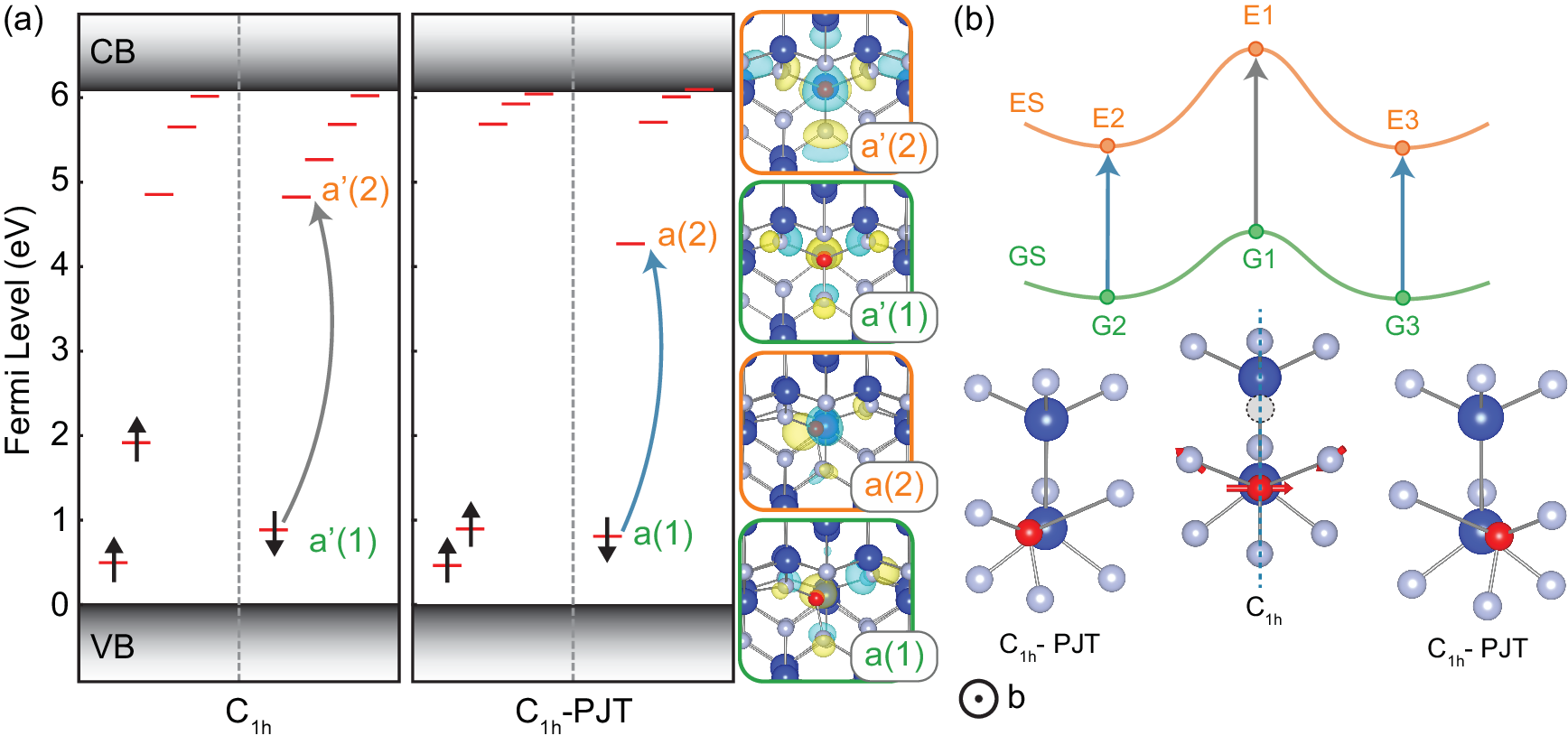}
    \caption{(a) Defect-level diagram of the NV$^{-}$ center in the C$_{1h}$ configuration (left) and the pseudo-Jahn-Teller (PJT) distorted configuration (C$_{1h}$-PJT, right). Energies are referenced to the valence band maximum (VBM). Red markers indicate Kohn-Sham levels, with a grey arrow denoting the transition from the occupied level [$a'$(1)] to the unoccupied level [$a'$(2)]. In the C$_{1h}$-PJT configuration, the localized KS levels are labeled as $a(1)$ and $a(2)$ due to the absence of symmetry elements, and a blue arrow indicates the transition between them. Isosurfaces of the real part (30\% of the maximum) of the wave functions corresponding to occupied and unoccupied levels in the minority-spin channel are shown. (b) Schematic potential energy surfaces (PES) for the ground state (GS, green) and excited state (ES, orange). The PES nodes represent the symmetric ground state (G1), distorted ground states (G2 and G3), and corresponding excited states (E1, E2, and E3). The phonon mode (red arrows) responsible for the distortion from C$_{1h}$ (G1) to C$_{1h}$-PJT configurations (G2 and G3) is illustrated. The energy differences between configurations are: $\Delta E_{\text{G1-G2}} = 1.00$ eV, $\Delta E_{\text{E1-E2}} = 1.66$ eV, $\Delta E_{\text{E1-G1}} = 2.46$ eV, and $\Delta E_{\text{E2-G2}} = 1.80$ eV. The changes in configuration coordinate are: $\Delta Q_{\text{G1-G2}} = 2.53$ amu$^{1/2}$\AA{}, $\Delta Q_{\text{E1-E2}} = 2.63$ amu$^{1/2}$\AA{}, $\Delta Q_{\text{E1-G1}} = 0.44$ amu$^{1/2}$\AA{}, and $\Delta Q_{\text{E2-G2}} = 0.36$ amu$^{1/2}$\AA{}.}
    \label{figure2}
\end{figure*}

Nitrogen vacancy introduces three planar dangling bonds derived from the 3\textit{s} and 3\textit{p} orbitals of the neighboring Si atoms. Their interaction produces localized Kohn-Sham (KS) levels within the bandgap (see Fig. S2 in the SI). When a nitrogen atom substitutes a nearby Si site to form the N$_\text{Si}$V$_\text{N}$ center, hybridization between the substituted N and surrounding N atoms alters the KS levels. The negatively charged defect features an occupied KS level $a'(1)$ just above the VBM and an unoccupied level $a'(2)$ located 3.94 eV higher in energy [see C$_{1h}$ configuration in Fig.~\ref{figure2}(a)]. In the NV$^{-}$ charge state, these two localized states transform as $A'$ irreducible representations, denoted as $a'(1)$ (occupied) and $a'(2)$ (unoccupied). Both the ground and excited states are spin doublets and represented as $^2A'$. Optical transitions between them are linearly polarized in the $xy$-plane, producing a ZPL at 2.46 eV with a transition dipole moment (see Eq. S16 in SI) of 4.7 Debye. This is in close agreement with the experimentally reported ZPL near 2.26 eV~\cite{Senichev2021, Martin2023}. The corresponding radiative rate and lifetime (see Eq. S15 in SI) are 111 MHz and 9.01 ns, respectively. Our calculated values also agree well with ZPL ($\sim$2.15 eV) and lifetime (3$-$8 ns) reported by Chen \textit{et al.}~\cite{Chen2025}. Similar nitrogen antisite-vacancy complex defect has also been observed as source of $\sim$2 eV SPE in GaN~\cite{Yuan2023}. In the majority-spin channel [Fig.~\ref{figure2} (a)], the occupied $a'$ and unoccupied $a''$ states give rise to a second dipole-allowed transition polarized along $z$, as only $\langle a' | z | a'' \rangle$ transforms as $A'$. This transition yields a ZPL at 0.85 eV, within the telecommunication spectral range, with a transition dipole of 7.2 Debye and a radiative rate and lifetime of 10.5 MHz and 95.24 ns, respectively.

The NV$^{-}$ center in the C$_{1h}$ configuration exhibits an imaginary phonon mode. This mode corresponds to a localized out-of-plane oscillation of the substituted nitrogen (N$_{Si}$) between neighboring nitrogen atoms [Fig.~\ref{figure2}(b)], indicating that the C$_{1h}$ configuration is an unstable saddle point in the potential energy surface. It originates from the shorter N-N bond compared to the Si-N bond. A small displacement along this mode drives the system toward a lower-symmetry configuration, in which the substituted nitrogen moves closer to one neighboring nitrogen, breaking the mirror symmetry [see Fig.~\ref{figure2}(b) and Fig. S7 in SI]. Since the C$_{1h}$ configuration is nondegenerate yet still distorts, this behavior is classified as a pseudo-Jahn-Teller (PJT) effect~\cite{Bersuker2006}. This distortion results in C$_1$ point group symmetry and is denoted as C$_{1h}$-PJT from now onwards [Fig.~\ref{figure2}(b)]. The two possible distortions, toward either neighboring nitrogen, are energetically degenerate, with approximately 1~eV lower energy than the high-symmetry C$_{1h}$ structure. The distortion of the C$_{1h}$ configuration of the NV$^-$ defect modifies the energy and symmetry of the localized states [see Fig.~\ref{figure2}(a)], altering its optical properties. In the minority-spin channel, C$_{1h}$-PJT configuration exhibits a ZPL at 1.80~eV and a radiative lifetime of 10.17~ns, in close agreement with experimental reports~\cite{Senichev2021}. In realistic samples, the NV$^-$ center may experience diverse local environments, such as those near interfaces or regions under strain, that can alter its electronic and optical characteristics. Given that the optical response is sensitive to subtle changes in the atomic configuration, as illustrated by the differences in properties of the C$_{1h}$ and C$_{1h}$-PJT structures, such local variations likely contribute to the spread and variability observed in experimental measurements.

\begin{figure*}
    \centering
    \includegraphics[width=\linewidth]{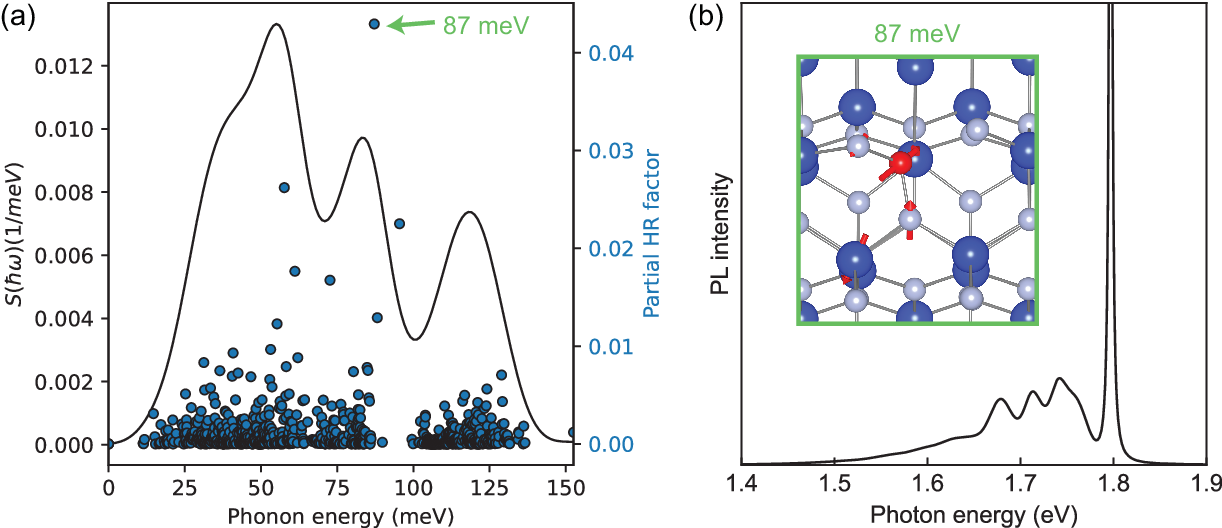}
    \caption{(a) Spectral function [$S(\hbar\omega)$] (black line, left axis) and partial Huang-Rhys factors (blue dots, right axis) with phonon energy, illustrating the energy-resolved electron-phonon coupling and dominant phonon modes associated with the $a(2) \rightarrow a(1)$ electronic transition in the C$_{1h}$-PJT configuration of the NV$^{-}$ defect. (b) Calculated photoluminescence (PL) spectrum showing a pronounced ZPL emission at 1.80 eV and the vibronic sideband features, highlighting the role of electron-phonon coupling in the optical emission profile. The inset illustrates the dominant phonon mode at 87 meV.}
    \label{figure3}
\end{figure*} 

Next, we analyzed the electron-phonon coupling of this transition using the method of Alkauskas \textit{et al.}~\cite{Alkauskas2014}. The photoluminescence spectra show a phonon sideband with three peaks, also reflected in the calculated electron-phonon spectral function [Fig.~\ref{figure3}]. The 87 meV phonon mode with the largest partial Huang-Rhys (HR) factor is quasilocalized with a localization ratio of 14.4, while the spectral peaks arise from multiple phonon modes with different degrees of localization [see Fig.~\ref{figure3} and Fig. S3 in SI]. The first peak near 55~meV agrees with the first peak of phonon sideband in the range 38.6-57.4~meV observed in experiment~\cite{Chen2025}, with additional peaks at 83 and 118~meV. The total HR factor is 0.9, corresponding to a DW factor of 41$\%$, which is close to the experimentally measured range of 50-70$\%$ reported by Chen \textit{et al.}~\cite{Chen2025}. These DW values are significantly higher than the DW factor of 3$\%$ for the well-known NV$^-$ center in diamond, even though the ZPL energy and excited-state lifetime are comparable, at approximately $\sim 2$ eV and 12~ns~\cite{Batalov2008, Gali2019}. The moderate electron-phonon interaction (HR factor = 0.9) ensures that a large fraction of the optical emission remains in the zero-phonon line [see Fig.~\ref{figure3}(b)]. This behavior arises because the structural relaxation between the ground and excited states is relatively small, with a change in configuration coordinate [$\Delta$Q, see Eq. S11 in SI] of only 0.36~amu$^{1/2}$\AA{}. For the C$_{1h}$ configuration, the HR and DW factors are 1.11 and 33$\%$, respectively. The strength of the electron-phonon coupling, expressed by the HR factor, is reflected in the calculated $\Delta$Q for both the C$_{1h}$ (0.44~amu$^{1/2}$\AA{}) and C$_{1h}$-PJT (0.36~amu$^{1/2}$\AA{}) geometries. We note that PBE functional underestimates the ground-to-excited state configurational displacement $\Delta$Q relative to that calculated from HSE functional~\cite{razinkovas2021vibrational}, leading to reduced HR factor and overestimated Debye-Waller values. While SCAN provides an improvement over PBE, it still underestimates experimental HR values for the NV$^{-}$ center in diamond~\cite{maciaszek2023application}. Thus, our reported HR values represent a reliable lower bound. Earlier studies~\cite{Senichev2021, Martin2023} proposed that the observed photoluminescence features might originate from a single family of defects with a complex structure. Our results support this interpretation and indicate that the experimentally observed emission is consistent with nitrogen-vacancy-related centers in silicon nitride, rather than defects in a silica host~\cite{Chen2025}. Further controlled synthesis and targeted experiments will be essential to verify the microscopic origin of these single-photon emitters. Our findings will guide future efforts to isolate and engineer bright quantum emitters in silicon nitride.

In summary, we identify nitrogen-vacancy-related defects, particularly NV$^{-}$, as the likely origin of the recently observed $\sim$2~eV single-photon emission in silicon nitride. The NV$^{-}$ exhibits a linearly polarized ZPL emission at 2.46~eV in the C$_{1h}$ configuration, with a lifetime of 9.01 ns and a high DW factor of 33\%. We further show that pseudo-Jahn-Teller distortions generate symmetrically equivalent configurations (C$_{1h}$-PJT), which modify the electronic structure and emission characteristics with a ZPL, lifetime, and DW factor of 1.80 eV, 10.17 ns, and 41\%, respectively. The moderate electron-phonon coupling during the electronic transition gives a higher emission in the ZPL compared to the phonon-side band. Our study provides a microscopic understanding of bright quantum emitters and suggests that controlled creation of nitrogen-vacancy-related defects could enable deterministic integration of single-photon sources in silicon-nitride photonic circuits. This work lays the foundation for the design and implementation of scalable quantum photonic devices based on silicon nitride.

The authors thank the Materials Research Centre (MRC), the Supercomputer Education and Research Centre (SERC), and the Solid State and Structural Chemistry Unit (SSCU) at the Indian Institute of Science, Bangalore, for access to computing resources. This work was supported by the DST-Nanomission program of the Department of Science and Technology, Government of India (Grant No. DST/NM/TUE/QM-1/2019). S.M. acknowledges funding from the PMRF fellowship (ID No. 0201908). The authors also acknowledge support from the Institute of Eminence (IoE) scheme of the Ministry of Human Resource Development, Government of India.

\providecommand{\latin}[1]{#1}
\makeatletter
\providecommand{\doi}
  {\begingroup\let\do\@makeother\dospecials
  \catcode`\{=1 \catcode`\}=2 \doi@aux}
\providecommand{\doi@aux}[1]{\endgroup\texttt{#1}}
\makeatother
\providecommand*\mcitethebibliography{\thebibliography}
\csname @ifundefined\endcsname{endmcitethebibliography}  {\let\endmcitethebibliography\endthebibliography}{}

\clearpage
\widetext
\begin{center}
\textbf{\large Supplemental Materials: Origin of Bright Quantum Emissions with High Debye-Waller factor in Silicon Nitride}
\end{center}
\setcounter{equation}{0}
\setcounter{figure}{0}
\setcounter{table}{0}
\setcounter{page}{1}
\makeatletter
\renewcommand{\theequation}{S\arabic{equation}}
\renewcommand{\thefigure}{S\arabic{figure}}
\renewcommand{\bibnumfmt}[1]{[S#1]}
\renewcommand{\citenumfont}[1]{S#1}

\section{Section S1: Bulk Properties}

The bandgaps of $\beta$-Si$_3$N$_4$ obtained from various theoretical approaches and experimental measurements are summarized in Table~\ref{table_bg_transposed}. Our calculated bandgap using hybrid functional parameters \texttt{AEXX} = 0.3 and \texttt{HFSCREEN} = 0.2 closely matches both the $G_0W_0$ computed value and the experimentally reported bandgap of 6.5~eV for crystalline $\beta$-Si$_3$N$_4$, as measured by linear fitting of EELS-STEM data by Asano et al.~\cite{Asano2022}. However, discrepancies persist among experimentally reported bandgap values in the literature.

\begin{table}[h]
    \centering
    \begin{tabular}{|c|c|c|c|c|c|c|}
        \hline
        Method & PBE & DDH & HSE06 & HSE$^a$ & G$_0$W$_0$$^b$ & Exp.$^c$ \\
        \hline
        E$_g$ (eV) & 4.4 & 5.8 & 5.8 & 6.1 & 6.2 & 6.5 \\
        \hline
    \end{tabular}
    
    \vspace{0.2cm}
    \begin{minipage}{\textwidth}
    \small
    $^a$With AEXX=0.3 and HFSCREEN=0.2 \\
    $^b$GW oneshot calculation \\
    $^c$Experimental bandgap from linear fit in EELS-STEM~\cite{Asano2022}
    \end{minipage}
    \caption{Bandgap of $\beta$-Si$_3$N$_4$ obtained from various computational methods and experimental measurement.}
    \label{table_bg_transposed}
\end{table}

\section{Section S2: Theoretical Methodology}

\subsubsection{\textbf{S2.1: Defect Formation Energy}}

The defect formation energy \( E^{f}[X^{q}] \) of a defect \( X \) in charge state \( q \), as a function of the Fermi level \( E_F \), is calculated using the expression~\cite{Zhang1991,Kumagai2014}:

\begin{equation}
E^{f}[X^{q}] = E_{\text{tot}}[X^{q}] - E_{\text{tot}}[\text{bulk}] - \sum_i n_i \mu_i + qE_F + E_{\text{corr}}
\end{equation}

Here, \( E_{\text{tot}}[X^{q}] \) and \( E_{\text{tot}}[\text{bulk}] \) are the total energies of the defect-containing and pristine bulk supercells, respectively. The term \( \sum_i n_i \mu_i \) accounts for the exchange of atoms with reservoirs, where \( n_i \) is the number of atoms of species \( i \) added (\( n_i > 0 \)) or removed (\( n_i < 0 \)), and \( \mu_i \) is the corresponding chemical potential.

The chemical potentials are defined as follows:
\begin{itemize}
    \item Under N-rich conditions: \( \mu_N^{\text{N-rich}} = \frac{1}{2}E_{\text{tot}}(\text{N}_2) \)
    \item Under Si-rich conditions: \( \mu_{Si}^{\text{Si-rich}} = E_{\text{tot}}(\text{Si}) \)
\end{itemize}

The remaining chemical potentials are derived from the constraint:
\begin{equation}
3\mu_{Si} + 4\mu_N = \mu_{Si_3N_4}
\end{equation}
where \( \mu_{Si_3N_4} \) is the total energy per formula unit of crystalline Si$_3$N$_4$.

The term \( qE_F \) represents the energy required to exchange electrons with the reservoir defined by the Fermi level \( E_F \). \( E_{\text{corr}} \) includes finite-size corrections for potential alignment and charged defect interactions. We employ the extended Freysoldt-Neugebauer-Van de Walle (eFNV) correction as implemented in the \texttt{DOPED} package~\cite{Kumagai2014,Kavanagh2024}.

\subsubsection{\textbf{S2.2: $\Delta$SCF Calculation}}
For the C$_\text{1h}$ configuration, the ground and excited states are both ${}^2A'$ (doublets), while for the C$_\text{1h}$-PJT distorted structure (C$_1$ symmetry), those are both ${}^2A$. The ground and excited states of the C$_\text{1h}$ configuration can be represented by single Slater determinants as 
\[
\left| a'(1)\,\overline{a'(1)}\,a'(2) \right\rangle 
\quad \text{and} \quad 
\left| a'(1)\,\overline{a'(2)}\,a'(2) \right\rangle,
\]
respectively, where the overbar denotes electron occupation in the minority-spin channel. 

Similarly, in the C$_\text{1h}$-PJT configuration, the ground and excited states can be represented by single Slater determinants as
\[
\left| a(1)\,\overline{a(1)}\,a(2) \right\rangle 
\quad \text{and} \quad 
\left| a(1)\,\overline{a(2)}\,a(2) \right\rangle,
\]
respectively. Therefore, the $\Delta$SCF approach is applicable here because these correspond to doublet-to-doublet transitions dominated by single Slater determinants.

\subsubsection{\textbf{S2.3: Photoluminescence and Electron-Phonon Coupling}}

To study the electron-phonon coupling and photoluminescence (PL) spectrum for defect-to-defect transitions, we use the theory of Huang and Rhys~\cite{HR1950}, as demonstrated by Alkauskas et al.~\cite{alkauskas2012first} and implemented in custom built \texttt{DEFECTPL} package~\cite{defectpl}. The normalized PL lineshape is given by:

\begin{equation}
L(\hbar\omega) = C\omega^3A(\hbar\omega)
\end{equation}

where the optical spectral function \( A(\hbar\omega) \) is defined as:

\begin{equation}
A(\hbar\omega) = \sum_{m}{|\langle \chi_{gm} | \chi_{e0} \rangle|^2\delta (E_{ZPL} - E_{gm} - \hbar\omega)}
\end{equation}

Here, \( \chi_{gm} \) and \( \chi_{e0} \) are vibrational wavefunctions of the ground and excited states, respectively, and \( E_{gm} = \sum_k n_k\hbar\omega_k \) is the energy of the ground-state vibrational level.

Assuming similar phonon modes in both states and using the generating function approach~\cite{Lax1952,Kubo1955}, the spectral function becomes:

\begin{equation}
A(E_{ZPL} - \hbar\omega) = \frac{1}{2\pi}\int_{-\infty}^{\infty}{G(t)e^{i\omega t - \gamma |t|}dt}
\end{equation}

where \( G(t) = e^{S(t)-S(0)} \) is the generating function and \( \gamma \) determines the ZPL width. The function \( S(t) \) is the Fourier transform of the electron-phonon coupling spectral function \( S(\hbar\omega) \):

\begin{equation}
S(t) = \int_{0}^{\infty}{S(\hbar\omega)e^{-i\omega t}d(\hbar \omega)}
\end{equation}

The total Huang-Rhys factor is:

\begin{equation}
S(0) = \int_{0}^{\infty}{S(\hbar\omega)d(\hbar\omega)} = \sum_k S_k
\end{equation}

Each partial HR factor \( S_k \) is given by:

\begin{equation}
S_k = \frac{\omega_kq_k^2}{2\hbar}
\end{equation}

with:

\begin{equation}
q_k = \sum_{\alpha,i}{\sqrt{m_\alpha}(R_{\alpha,i}^{(e)}-R_{\alpha,i}^{(g)})\Delta r_{k,\alpha,i}}
\end{equation}

Here, \( m_\alpha \) is the mass of atom \( \alpha \), \( R_{\alpha,i}^{(g)} \) and \( R_{\alpha,i}^{(e)} \) are atomic coordinates in the ground and excited states, and \( \Delta r_{k,\alpha,i} \) is the component of the normalized displacement vector of the \( k^{\text{th}} \) phonon mode.

The spectral function is then:

\begin{equation}
S(\hbar\omega) = \sum_k S_k \delta(\hbar\omega - \hbar\omega_k)
\end{equation}

The total HR factor \( S(0) \) quantifies the electron-phonon coupling and is reflected in the configuration coordinate change:

\begin{equation}
\label{del_q}
\Delta Q = \sqrt{\sum_{\alpha} m_\alpha (\Delta R_\alpha)^2}
\end{equation}

where \( \Delta R_\alpha \) is the difference in atomic coordinates of atom \( \alpha \) between ground and excited states.

\subsubsection{\textbf{S2.4: Debye-Waller Factor and Localization Ratio}}

The Debye-Waller factor represents the fraction of emission in the ZPL, is:

\begin{equation}
w_{ZPL} = e^{-S(0)}
\end{equation}

Phonon mode localization is estimated using the localization ratio:

\begin{equation}
\beta_k = \frac{N}{IPR_k}
\end{equation}

where \( N \) is the number of atoms in the supercell and \( IPR_k \) is the inverse participation ratio~\cite{alkauskas2012first}:

\begin{equation}
IPR_k = \frac{1}{\sum_{\alpha}{\left(\sum_i \Delta r_{k,\alpha,i}^2\right)^2}}
\end{equation}

\subsubsection{\textbf{S2.5: Radiative Transition Rate and Lifetime}}

The radiative transition rate \( \Gamma_{\text{rad}} \) and lifetime \( \tau_{\text{rad}} \) are calculated as:

\begin{equation}
\Gamma_{\text{rad}} = \frac{1}{\tau_{\text{rad}}} = \frac{n_rE_{ZPL}^3|\Bar{\mu}|^2}{3\pi\epsilon_0c^3\hbar^4}
\end{equation}

Here, \( n_r = 2.0458 \) is the refractive index of $\beta$-Si$_3$N$_4$, and \( |\Bar{\mu}| \) is the magnitude of the transition dipole moment in e\AA. The constants \( \epsilon_0 \), \( c \), and \( \hbar \) are used in units of \( \frac{e^2}{\text{eV.\AA}} \), \AA/s, and eV·s, respectively.

The transition dipole moment between ground and excited single-particle states is computed using the \texttt{VaspBandUnfolding} package~\cite{qijing}:

\begin{equation}
\mathbf{\mu}_k = \frac{i\hbar}{(\epsilon_{f,k}-\epsilon_{i,k})m}\langle \psi_{i,k} | \mathbf{p} | \psi_{f,k} \rangle
\end{equation}

\clearpage
\section{Section S3: Defect Thermodynamics in Si-rich Conditions}

Under Si-rich conditions, the nitrogen vacancy is found to be the most stable defect with a maximum formation energy reaching up to 4 eV for the neutral charge state. However, nitrogen substitution at silicon sites becomes energetically unfavorable with higher formation energies, resulting in less stable antisite and antisite-vacancy defects involving nitrogen.

\begin{figure}[h]
    \centering
    \includegraphics[width=0.5\linewidth]{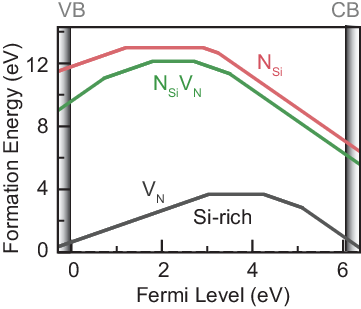}
    \caption{Formation energy of nitrogen vacancy, nitrogen antisite, and nitrogen antisite-vacancy defects in Si-rich growth conditions.}
    \label{fig:formation_energy}
\end{figure}

\clearpage
\section{Section S4: Electronic Structure of Neutral Nitrogen Vacancy}

\begin{figure}[ht]
    \centering
    \includegraphics[width=0.8\linewidth]{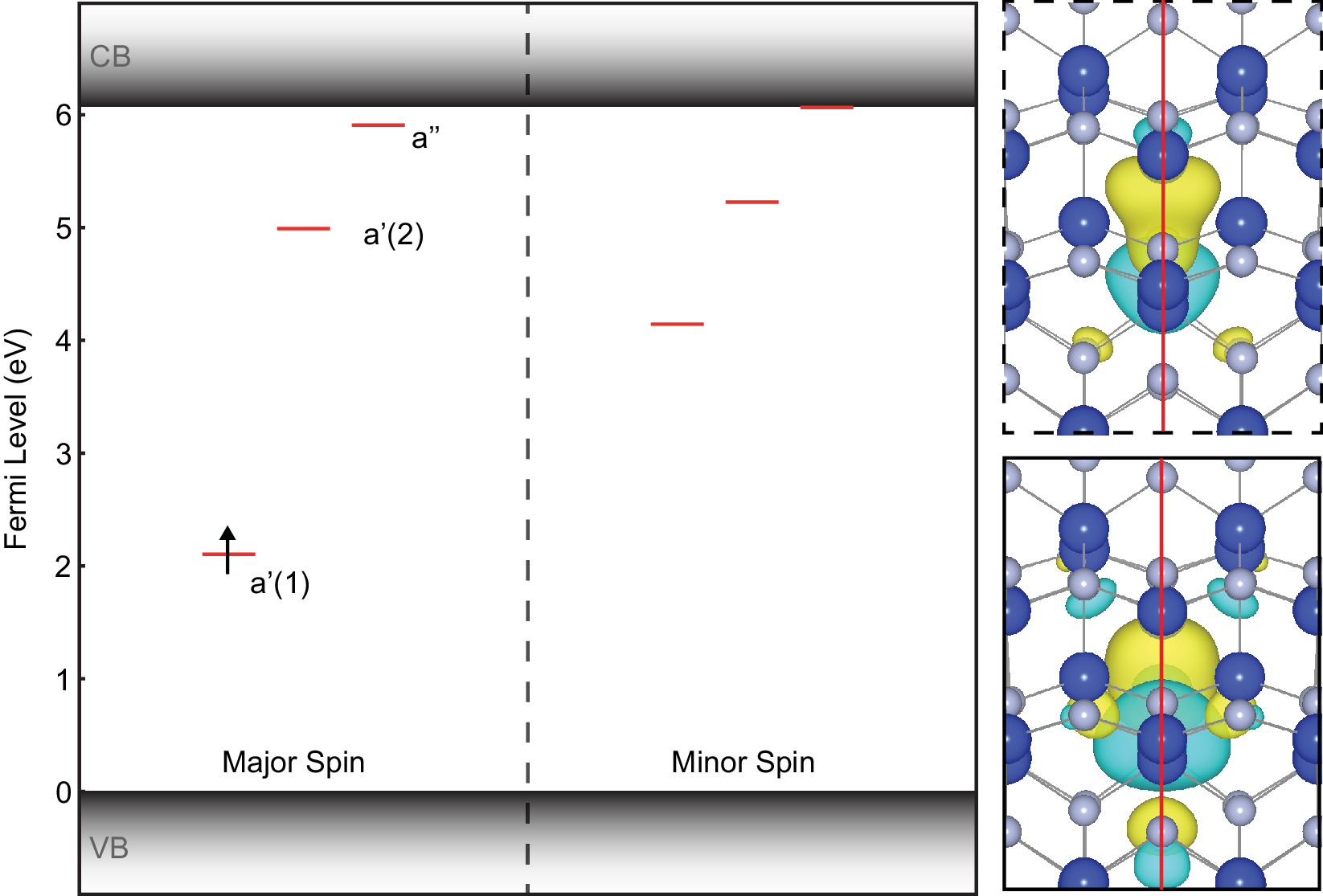}
    \caption{Localized states formed inside the bandgap of $\beta$-Si$_3$N$_4$. The left panel shows Kohn-Sham levels of neutral V$_\text{N}$ defects in both spin channels within the bandgap. The Fermi level is shown on the vertical axis with values referenced to the valence band maximum. The irreducible representations of localized KS levels are indicated according to the C$_{1h}$ point group symmetry of the defect. The right panel shows the real part of the wave function of the highest occupied (solid border) and lowest unoccupied (dashed border) localized levels. The red line represents the mirror plane symmetry element perpendicular to the c-axis.}
    \label{fig:localized_states}
\end{figure}

\clearpage
\section{Section S5: Localization Ratio of Phonon Modes}

\begin{figure}[h]
    \centering
    \includegraphics[width=0.6\linewidth]{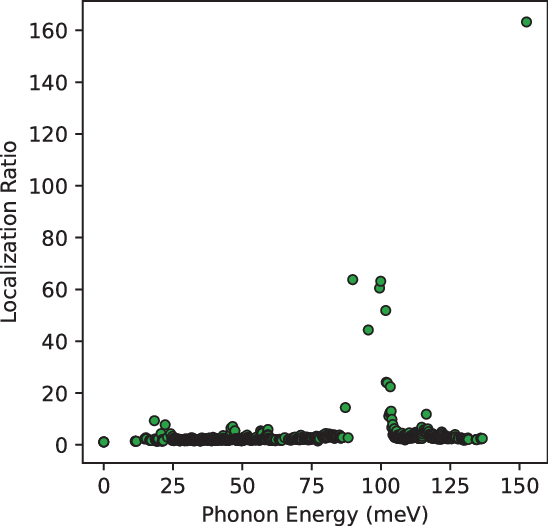}
    \caption{Localization ratio of phonon modes for NV$^{-}$ defect in C$_{1h}$-PJT configuration.}
    \label{fig:localization_ratio}
\end{figure}

\section{Section S6: Migration Barrier}
\begin{figure}[h]
    \centering
    \includegraphics[width=0.6\linewidth]{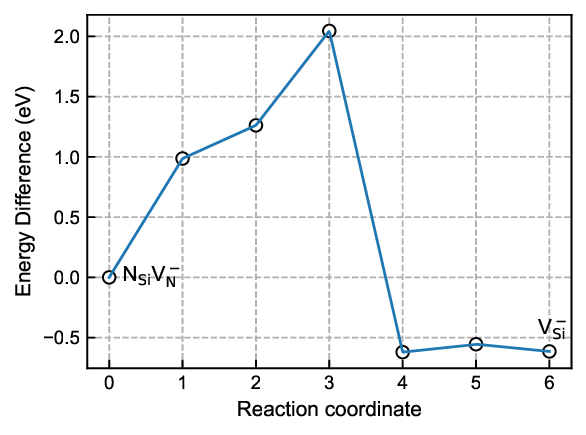}
    \caption{Migration barrier from N$_{Si}$V$_N^-$ to V$_{Si}^-$ and vice-versa.}
    \label{fig:neb}
\end{figure}
We performed nudged elastic band (NEB) calculation to evaluate the feasibility of a neighbouring nitrogen atom migrating from the nitrogen site adjacent to a silicon vacancy to the silicon vacancy site. As shown in Figure~\ref{fig:neb}, our results indicate that this process involves a substantial energy barrier of 2.66 eV, as shown below. Furthermore, once the nitrogen atom is at the silicon site, the reverse transition back to the nitrogen site is also highly unlikely, with a calculated barrier of 2.04 eV.

\section{Section S7: Local Geometry}
\begin{figure}[h]
    \centering
    \includegraphics[width=0.6\linewidth]{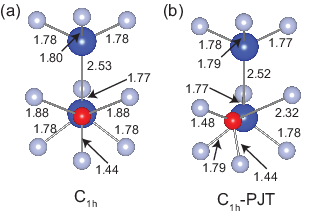}
    \caption{Bond lengths (Si–N, N–N, and Si–Si) around the N$_\mathrm{Si}$V$_\mathrm{N}^-$ defect in 
(a) the C$_{1h}$ configuration and (b) the C$_{1h}$-PJT configuration. 
All bond lengths are given in angstrom (\AA{}). The substituted nitrogen atom (shown in red) 
displaces by 0.37\,\AA{} during the distortion.}
    \label{fig:bond_analysis}
\end{figure}

The bond lengths between atoms around N$_\mathrm{Si}$V$_\mathrm{N}^-$ defect in 
the C$_{1h}$ configuration and the C$_{1h}$-PJT configuration are shown in Figure~\ref{fig:bond_analysis}.

\clearpage

\providecommand{\latin}[1]{#1}
\makeatletter
\providecommand{\doi}
  {\begingroup\let\do\@makeother\dospecials
  \catcode`\{=1 \catcode`\}=2 \doi@aux}
\providecommand{\doi@aux}[1]{\endgroup\texttt{#1}}
\makeatother
\providecommand*\mcitethebibliography{\thebibliography}
\csname @ifundefined\endcsname{endmcitethebibliography}  {\let\endmcitethebibliography\endthebibliography}{}

\end{document}